# Wavefront Shaping of Ultrasound Vortex through the Human Skull Enabled by Binary Acoustic Metasurfaces

Zhongtao Hu[1,2,3*]

1. School of Engineering Medicine, Beihang University, Beijing 100191, China
2. Beijing Advanced Innovation Center for Biomedical Engineering, Beijing 100191, China.
3. Shenzhen Institute of Beihang University, Shenzhen 518000, China.

Address correspondence to Zhongtao Hu, Ph.D. School of Engineering Medicine, Beihang University. 37 Xueyuan Road, Haidian District, Beijing 100191, China.

[*]Corresponding author: Zhongtao Hu (zhongtaohu@buaa.edu.cn)

**Abstract:**

Ultrasound vortices have rapidly expanded their applications to areas like particle trapping, contactless manipulation, acoustic communications. In ultrasonic imaging and therapy involving bone tissues, these vortex beams offer intriguing possibilities but transmitting them through bone (especially the skull) poses challenges. Traditional acoustic lenses were engineered to rectify skull-induced beam aberration, and their capacity was limited to generating only static ultrasound fields within the brain. To overcome this constraint, our study presents a novel method for transcranially steering focused ultrasound vortex using 3D printed binary acoustic metasurfaces (BAMs) with a thickness of 0.8 λ. We tackled the challenge of skull-induced phase aberration by computing the phase distribution via a time reversal technique, which concurrently enabled the generation of a steerable focused vortex inside an *ex vivo* human skull by adjusting the operating frequency. Both numerical and simulations experiments were conducted to validate the capabilities of BAMs. Furthermore, we explored the generation of higher-order topological charge acoustic vortices within the brain utilizing the BAM. This development paves the way for designing cost-effective particle-trapping systems, facilitating clot manipulation, and applying acoustic-radiation forces and torques within or across bone structures, thus presenting a new frontier for potential biomedical applications.

**Keywords:** Transcranial Focused Ultrasound, Acoustic Vortex Beam, Skull-Induced Phase Aberration Correction, Ultrasound Beam Steering Through Bone, Binary Acoustic Metasurface

## 1. Introduction

The development of specialized wavefront-shaping techniques to achieve reliable focusing through bone structures is a critical consideration for both ultrasonic diagnostic imaging and therapeutic applications[1–6]. One promising advance in ultrasonic wavefront engineering is the use of acoustic vortex beams. Acoustic vortex beams characterized by their unique rotating wavefronts formed by intertwined helices, offer an unparalleled capability to remotely manipulate particles[7–11]. The inception of these beams can be traced back to the domain of optical manipulation, a groundbreaking concept that received accolades in the form of the 2018 Nobel Prize in Physics[12], and was later translated to acoustics by Hefner and Marston's pioneering work on helicoidal ultrasound transducers [13]. It holds transformative potential in fields like physics and biology[10,14–21]. Evidence of such potential can be seen in pioneering experiments, such as the manipulation of glass spheres in a pig bladder[22], the trapping of microbubbles in mouse back epidermal blood vessels[23], the manipulation of microparticles in zebrafish embryos[24] and acoustic manipulation of GV-expressing bacteria[25].

Recent innovations highlight the advantages of focused acoustic vortices. Their capacity for a stronger trapping force, the potential for deeper tissue penetration, and their ability to exert more significant torque make them highly promising [26]. In parallel, the emergence of transcranial focused ultrasound (tFUS) techniques has been noteworthy. This non-invasive method, capable of modulating neural activity[27–30], targeted drug delivery[31–36], and even ablating pathological tissues[37–41], stands poised to usher in a new era in medical treatment and research. The integration of acoustic vortex beams with tFUS is especially compelling: it allows imparting both linear momentum (from focusing) and angular momentum (from vorticity) to targets inside the brain[42–44]. This could refine the manipulation of particles, cells, or drug carriers within bone-encased structures and provide new ways to stimulate or modulate neural circuits with mechanical torque in addition to pressure.

Despite this promise, significant challenges remain for delivering focused vortex ultrasound through bone. The human skull is a thick irregular layer of cortical bone with an inner trabecular (diploë) core; its complex geometry, stiffness, and poroelastic structure cause strong scattering, refraction, and attenuation of ultrasound. This often results in ultrasound beam aberrations and defocusing, which also prevent the generation of acoustic vortex inside brain [45–48]. Although active device arrays with controllable phases offer a potential solution, these arrays need a large number of transducer elements and complex electronics [49–52]. Passive aberration-correcting elements like custom acoustic lenses or holograms have also been developed. Early 3D-printed acoustic lenses demonstrated that patient-specific phase correction is feasible by shaping a single-element transducer's wavefront according to skull CT scan data [53,54]. However, such conventional lenses produce only a fixed focus at one location (determined during fabrication). Once the lens is made and placed, it generates a static field optimized for a single target, unless the transducer or lens is mechanically moved. This static nature limits their utility for applications requiring dynamic steering or refocusing to different locations.

In light of these challenges, our study aims to develop a binary acoustic metasurface (BAM) lens that serves a dual purpose: (1) correct the wave distortions caused by the skull bone, and (2) enable dynamic steering of a focused ultrasound vortex transcranially (i.e., through the skull) without moving the device. By binarizing a holographic phase distribution, the BAM can manipulate the transmitted wavefront in a frequency-dependent manner. In essence, our design process combines the conjugate phase needed to counteract skull aberrations with the phase pattern required to generate an acoustic vortex. The resulting phase map is then reduced to two discrete phase levels (0 and $\pi/2$) to create a binary-phase metasurface that is simple to fabricate. We hypothesize that such a BAM, when placed between a single-element ultrasound transducer and the skull, can produce a high-fidelity focused vortex inside the brain and electronically steer its focal position along the beam axis by altering the driving frequency. In the following, we present the design methodology (including skull CT-based modeling and time-reversal phase computation), the fabrication of the BAM, and a comprehensive evaluation of its performance in correcting skull aberration and steering vortex beams. We emphasize the relevance of this approach to ultrasonic imaging and therapy of bone tissues – particularly transcranial applications – and discuss its potential clinical impact on bone-related ultrasound therapies and diagnostics.

## 2. Material and method

### 2.1 BAM design and fabrication

To design the BAM for transcranial focused ultrasound vortex (Fig. 1(a)), we followed a multi-step holographic approach. First, we extracted the geometry and acoustic properties of a human skull from CT images using empirical relations, and this provided a 3D acoustic model of the skull's heterogeneous structure for simulation purposes. Second, with the obtained skull properties, we performed numerical simulations using the time-reversal technique to calculate the acoustic wavefront generated by virtual point sources inside the skull (Fig. 1(b)). The acoustic wavefront was then recorded on the surface of the flat transducer located outside the skull and transformed into a phase profile as $P_0(x, y)$, the phase profile is compensated for the transcranial focusing of the flat source, which is same to our previous study[55]. Third, we designed a vortex-generating field whose phase is proportional to $P_a(x, y) = \exp(il\theta)$, where $\theta$ is the azimuthal angle and $l$ is the topological charge (Fig. 1(c)). The phase distribution of transcranial focused vortex given by the conjugation of skull-induced aberration corrections and the vortex-generating field as $P(x, y) = P_0(x, y)P_a(x, y)$ (Fig. 1(d)). Fourth, this $P(x, y)$ encodes both the corrections for skull-induced aberration and the helical phase for vortex generation. A continuous version of this phase map was initially obtained in the range $[0, 2\pi)$. We then binarized the phase map to two levels, 0 and $\pi/2$, by assigning each point on the aperture to the nearest of these two phase values with $\varphi(x, y) = 0$ for $p(x, y) > 0$ and $\varphi(x, y) = \pi/2$ for $p(x, y) < 0$ (Fig. 1(e)), and the binary phase map was rendered into a 3D-printed model, as shown in Fig. 1(f).

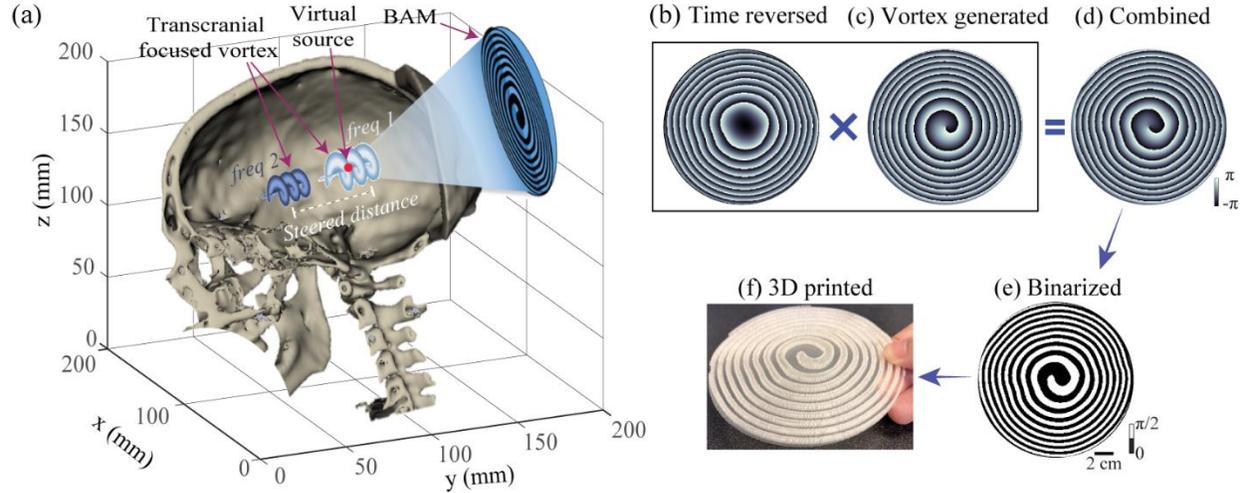

Fig. 1. Steering focused ultrasound vortex within the brain using BAM. (a) The illustration outlines the concept of transcranial steering of focused ultrasound vortex via BAM. Schematic representation of the BAM designed for transcranially steerable ultrasound vortex: (b) Time-reversal field derived from a virtual source waveform, (c) Vortex-generating field, (d) Combined phase distribution of both the Time-reversal and vortex-generating field, (e) Binarization of the combined field, and (f) Final 3D-printed BAM.

As a comparison, we also design the binary acoustic metasurface for the case of without skull. The binary design of BAMs make it work at broadband frequency range and enables the dynamic focusing properties as reported in previous studies [56,57], this is achievable due to the changed interaction between the binary acoustic lens and the varied wavelength, a consequence of altering the operating frequency of acoustic wave, which in turn modifies the resultant wavefront and the final focusing depth. However, it is difficult for the acoustic lens designed with gradient phase to achieve this property.

The designed BAM was then manufactured by 3D printing. The phase delay of each pixel on the BAM is proportional to the thickness of polylactic acid [58]. To produce a phase delay of $\pi/2$, the thickness ($d$) of unit '1' was calculated using $2\pi f d/c_1 - 2\pi f d/c_2 = \pi/2$, where $c_1$ and $c_2$ are the sound speed of the water and polylactic acid, respectively, and $f$ is the operating frequency. Thus, the thickness is represented as $d = c_1 c_2 / 4f(c_2 - c_1)$. The pressure transmission coefficient ($T$) of each unit '1' can be calculated using [59]: $T = \frac{2Z_r}{2Z_r \cos(2\pi f d/c_2) - i(Z_r^2+1)\sin(2\pi f d/c_2)}$, where the normalized acoustic impedance is given by $Z_r = Z_2/Z_1$, the impedance of water is given by $Z_1 = \rho_1 c_1$, and the impedance of polylactic acid is given by $Z_2 = \rho_2 c_2$. The terms $\rho_1$ and $\rho_2$ are the densities of water and polylactic acid, respectively. The acoustic properties of polylactic acid material are obtained experimentally using a pulse-echo technique in a cubic structure, resulting in a measured sound speed of 2212 m/s, and a density of 1223 kg/m³, and absorption of 3.54 dB/cm for 500 kHz. These measurements matched those reported in previous studies [60,61]. Water as the surrounding medium has a sound speed of 1480 m/s and mass density of 1000 kg/m³ at room temperature. The thickness of the BAM was calculated to be 2.35 mm for the operating frequency of 500 kHz, which was approximately 0.8λ (λ=2.97 mm). The BAM consisted of two printed parts. The first part was the polylactic acid unit on the lens with a depth of 2.25 mm that provided π/2 phase shift for the 500 kHz transmitted ultrasound wave. The second portion was the base plate printed at a thickness of 0.10 mm, which was needed to stabilize the lens. The transmission coefficient was 99.6% for unit '1', indicating efficient transmission through the lens.

## 2.2 Numerical and experimental method

The 3D-printed BAMs were evaluated numerically and experimentally using an *ex vivo* human skullcap. The *ex vivo* human skullcap, which was dry from storage in air, was immersed in water and degassed for a minimum of 24 h in a vacuum chamber at −0.1 MPa measured by a pressure gauge (Nisshin 1.6, Nisshin Seifun Group Inc., Tokyo, Japan) to eliminate air bubbles trapped in the skull before use.

For the numerical studies, simulations were performed using an open-source MATLAB toolbox, k-Wave, a pseudospectral method with k-space dispersion correction [62–65]. A graphics processing unit (Nvidia Tesla V100, Nvidia Corporation, Santa Clara, CA, USA) was used to accelerate the 3D simulations. The acoustic properties of the *ex vivo* human skullcap were obtained from CT scans using a clinical CT scanner (Siemens Somatom Confidence, Siemens Healthcare, Erlangen, Germany). The density and sound speed of the skull were converted from the Hounsfield units of the CT images by the function 'hounsfield2density' in the k-Wave toolbox. This function uses a piecewise linear fit to the experimental data reported by Schneider and Mast [66,67]. The density data of the skull ranges between $\rho_{min} = 1000$ kg/m³, and $\rho_{max} = 3200$ kg/m³, with an average density of $\rho_{mean} = 1525$ kg/m³, and the sound speed of skull ranges between $c_{min} = 1480$ m/s and $c_{max} = 4050$ m/s, with an average sound speed of $c_{mean} = 2542$ m/s, matching those reported in the previous literature [68,69]. The attenuation was obtained by the power law model as proposed by Aubry and Constans [70,71]. The size of each CT image was 512 × 512 × 351 voxels with a spatial resolution of 0.3906 mm in the xy-plane and 0.6000 mm in the z-axis. After linear interpolation, a numerical grid with an isotropic spatial resolution of 0.2 mm was generated. A numerical temporal step of $\Delta t = 20$ ns was used. The Courant-Friedrichs-Lewy number was

0.1 and the spatial sampling was approximately 15 grid points per wavelength in water at 500 kHz. These parameters were fixed for all simulations in this study.

For the experimental studies, the 3D-printed BAMs were coupled with a flat transducer as shown in Fig. 2(a). The flat transducer had a frequency of 500 kHz and an aperture of 120 mm. It was made of a single-element circular lead zirconate titanate (PZT) ceramic (DL-20, Del Piezo Specialties LLC, West Palm Beach, FL, USA). Positive and negative electrodes of the PZT ceramic were soldered with wires that were connected with an electrical driving system composed of a function generator (Model 33500B, Keysight Technologies Inc., Englewood, CO, USA) and a power amplifier (1020L, Electronics & Innovation, Rochester, NY, USA). The transducer was encased in a 3D-printed housing. The ultrasound waves generated by the flat transducer passing through a BAM and then skull was measured by a hydrophone (HGL-200, ONDA Corporation, Sunnyvale, CA) in a water tank filled with degassed and deionized water. The hydrophone was connected to a pre-amplifier (AG-20X0, Onda Corp., Sunnyvale, CA, USA) and a digital oscilloscope (Picoscope 5443D, St. Neots, United Kingdom) and moved in 3D using a computer-controlled 3D stage (PK245-01AA, Velmex Inc., NY, USA).

We validated both cases: (a) Without skull – using a homogeneous water medium to verify the BAM's vortex focusing design in ideal conditions, and (b) With skull – including the skull model to test aberration correction. Key output measures were the pressure field distribution in and beyond the focal region, which we analyzed for focal spot size, intensity patterns (to confirm the vortex's doughnut shape), and phase singularity at the vortex core. We also tested the BAM's frequency-dependent focusing by sweeping the source frequency from 450 kHz to 550 kHz to observe the shift in focal distance.

## 3. Results

### 3.1 BAM for focused ultrasound vortex without skull

In our pursuit of generating transcranial focused ultrasound vortices through BAMs, we initiated our investigation by designing, manufacturing, and evaluating BAMs for generating focused ultrasound vortex without skull. The designed and 3D printed BAMs, tailored for producing focused ultrasound vortex in homogeneous medium, are presented in Fig. 2(b). The magnitude of both simulated and experimentally measured acoustic pressure fields ($p(x, 0, z)/|p|$) in the xz-plane at y = 0 mm are depicted in Fig. 2(c). The data clearly demonstrates that the designed BAM effectively creates a beam with a void center, its pressure peaking around the focal distance at z = 55 mm. The pressure peaks at z = 55.0 mm in the simulation and z = 55.1 mm in the experiment. The simulated and measured transverse acoustic pressure fields in the xy-plan at the position of peak pressure are given in Fig. 2(d) and 2(e), respectively. The magnitude of fields shows a ring-shaped pattern with a central null. The null is caused by the phase singularity at the focal spot, which is clearly visible in Fig. 2(f) and 2(g) from the simulation and experiment, respectively. In addition, we can identify that the phase singularity is the result of a screw dislocation produced by the counterclockwise rotation of the phase, marked by the black arrows. In Fig. 2(h), the magnitude of the field is shown in detail for a transverse section measured at y = 0 and x = 2.0 mm, the numerical and experimental fields agree, and the beam shows a width of 4 mm (1.33 times of the wavelength at the nominal frequency) as shown in Fig. 2(i). Finally, the phase as a function of the azimuthal angle at a radial distance of 2.0 mm, corresponding to the half of the beam width, is shown in Fig. 2(j). The generated transverse field shows a levorotatory phase distribution whose phase rotates as a function of the azimuthal angle at a rate given by the topological charge, $l = 1$. Consequently, the phase undergoes a complete $2\pi$ radian rotation over an azimuthal turn, confirming that the designed BAMs for planar transducer can precisely generate acoustic vortices in a homogeneous medium.

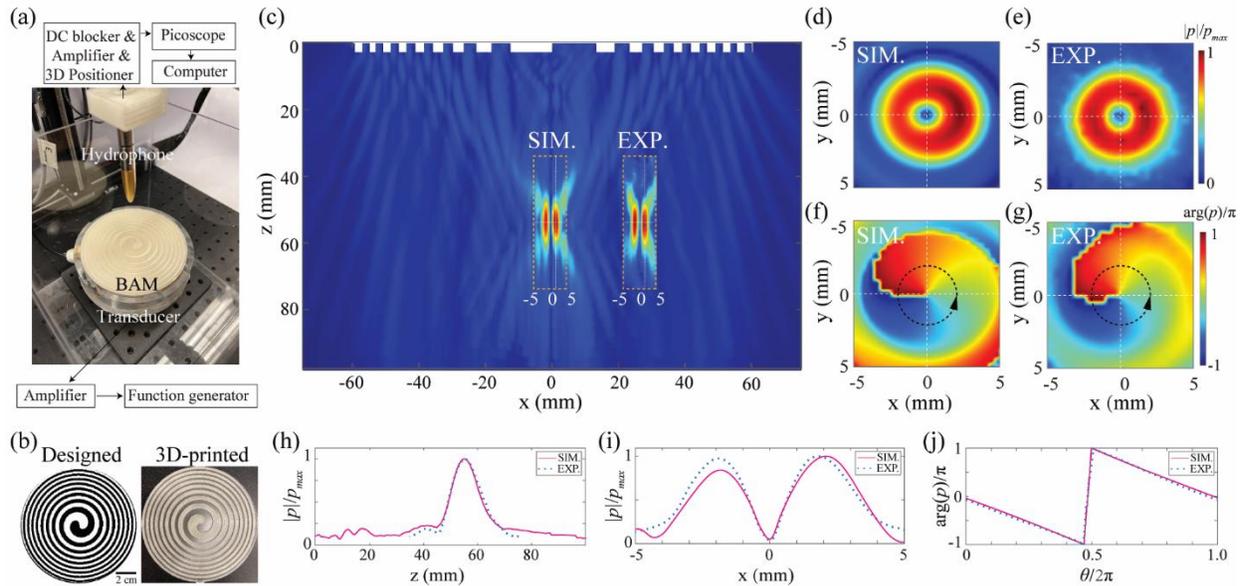

Fig. 2. Field of the focused ultrasound vortex (l = 1) designed for homogeneous medium. (a) Experimental setup for validating the designed BAM coupled with a plat ultrasound transducer. (b) The designed and 3D-printed BAMs without skull aberration correction. The white and black unit in the designed one represent units "$\pi/2$" and "0". (c) Simulated and experimental normalized sagittal cross section of the field at y = 0. (d), (e) Magnitude and (f), (g) phase of the transverse cross section of the field obtained at the position of peak pressure. (h) Axial cross section of the field measured at y = 0 and x = 2.0 mm. (i) Normalized magnitude of the field at z = 55 mm and y = 0. (j) Normalized phase of the field versus the azimuthal angle.

### 3.2 BAM for focused ultrasound vortex with skull

Then, we proceed to present the results for the BAM designed for focused ultrasound vortex through a skull, as summarized in Fig. 3, where the skull is placed on the surface of the BAM and the acoustic field is scanned across multiple planes using a hydrophone (Fig. 3(a)). To ensure precise alignment of the lens with the skull's surface, we use a 3D-printed skull holder. The holder's design takes into account the skull's irregular shape, guaranteeing that it can be correctly positioned only at the designated location. The designed and 3D printed BAM, tailored for producing focused ultrasound vortex through a skull, are presented in Fig. 3(b). The simulated and experimentally measured transcranial acoustic pressure fields in the xz-plane (y = 0 mm) generated by the BAM are displayed in Fig. 3(c).

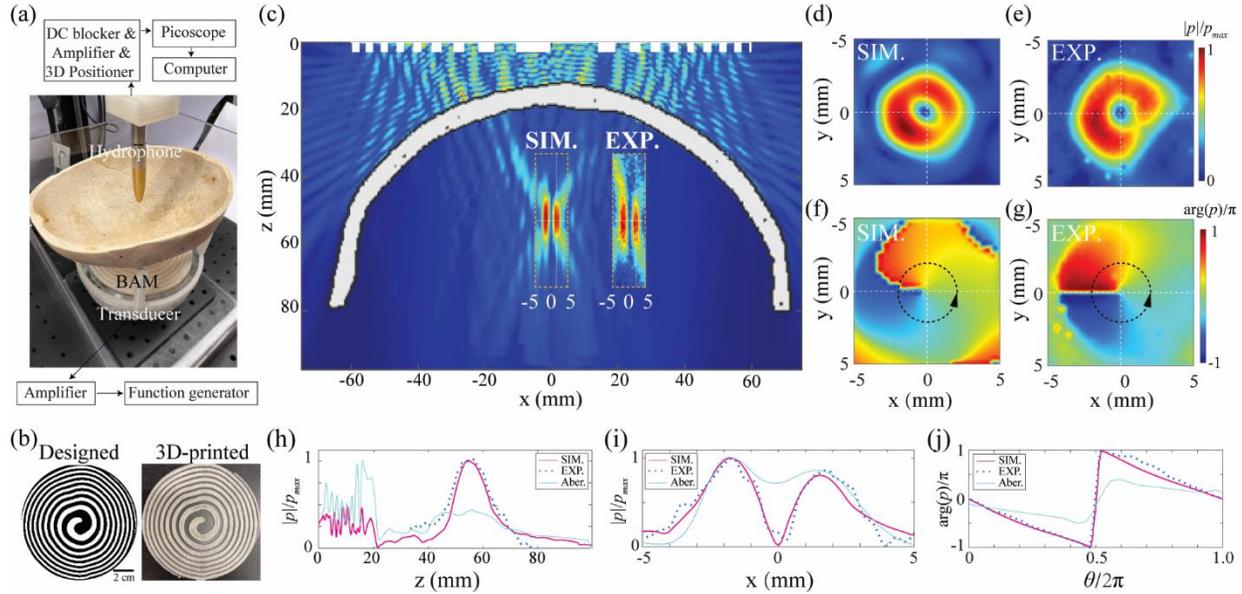

Fig. 3. Field of the ultrasound vortex ($l = 1$) designed for transcranial focusing. (a) Scheme of the experimental setup. (b) The designed and 3D-printed BAMs with skull aberration correction. The white and black unit in the designed one represent units "$\pi/2$" and "0". (c) Simulated and experimental normalized sagittal cross section of the field at y = 0. (d), (e) Magnitude and (f), (g) phase of the transverse cross section of the field obtained at the position of peak pressure obtained from simulated and experimental results. (h) Axial cross section of the field measured at y = 0 and x = 1.75 mm. (i) Normalized magnitude of the field at z = 55 mm and y = 0. (j) Normalized phase of the field versus the azimuthal angle. The light green line, labeled as (Aber.), represents the aberrated field resulting from transcranial focusing utilizing the lens designed for homogeneous medium.

The BAM is able to encode the time-reversed field with great accuracy. It compensates the aberrations of the skull while, simultaneously, it modifies the wavefront to generate a focused vortex by introducing a quasiparabolic phase profile with a rotational profile. Notably, both the simulated and experimental results closely align. The pressure peaks at z = 55.0 mm in the simulation and z = 55.3 mm in the experiment. The transverse cross section of the field at the position of peak pressure are shown in Figs. 3(d)–3(g). The transverse field magnitude [Figs. 3(d) and 3(e)] shows a ring-shaped distribution similar to that obtained for a homogeneous medium. The magnitude of the field vanishes at the center in both the experimental and the simulated results. The corresponding phase distribution is shown in Figs. 3(f) and 3(g). A counterclockwise rotation is achieved in the focal plane, and a phase singularity is visible at the location of the null. In Fig. 3(h), a detailed depiction of the field magnitude for a transverse section at y =

0 and x = 1.75 mm is presented. Both the numerical and experimental fields are in agreement, displaying a beam width of 3.5 mm (or 1.17 times the wavelength at the nominal frequency) as elaborated in Fig. 3(i). Finally, the phase as a function of the azimuthal angle at a radial distance of 1.75 mm, corresponding to the half of the beam width, is shown in Fig. 3(j). It's noteworthy that using a hologram designed for a homogeneous medium in transcranial applications leads to aberrations that distort the wavefront, as depicted in Figures 3(h-j). Contrarily, skull-induced aberrations distort the wavefront, reducing the focusing performance of the BAM compared to conditions in a homogeneous medium.

### 3.3 BAM for dynamic focusing of ultrasound vortex inside skull

A unique feature of the binary metasurface design is its ability to dynamically adjust the focal depth by changing the operating frequency of the incident ultrasound. We tested this capability transcranially by driving the same skull/BAM setup at different frequencies within the transducer's bandwidth. Figure 4 illustrates how the focus of the vortex beam can be electronically "steered" along the beam axis (the z-direction) without any physical movement. In both simulations and experiments, when the frequency was increased, the focal spot of the vortex moved closer to the transducer; when the frequency was decreased, the focus moved further away. Quantitatively, in the experiments the focal distance shifted from about 63.1 mm at 550 kHz to 48.2 mm at 450 kHz. At the mid-frequency of 500 kHz, as discussed, the focus was ~55 mm. This corresponds to a roughly linear relationship between frequency and focal depth: higher frequencies (shorter wavelengths) experience effectively a stronger "lens" action from the fixed phase pattern, focusing sooner, whereas lower frequencies (longer wavelengths) focus later. The linear fit of focal depth vs. frequency (Fig. 4(d)) from both simulation (solid squares) and experiment (open circles) confirms this trend. Importantly, the vortex nature of the beam remains intact across this frequency range. The transverse cross-sections at the new focal positions (e.g., Fig. 4(c) compares 450 kHz vs 550 kHz) still show the ring-shaped intensity distribution characteristic of the vortex. Thus, tuning frequency does not disrupt the orbital angular momentum content of the beam; it only repositions the focus. The practical range of this electronic steering is limited by the bandwidth of the transducer (in our case ~100 kHz span) and the fact that the binary phase design is optimized around a central frequency (performance degrades if going too far off-design). Nevertheless, within a reasonable bandwidth, our results demonstrate a continuous and reversible focal adjustment of nearly ±7 mm around the nominal focus. This dynamic focal shift was achieved without any mechanical translation of the transducer or lens, highlighting a key advantage of the BAM approach over traditional fixed lenses. Overall, these results confirm that the 3D-printed binary metasurface can electronically steer a transcranial focused ultrasound vortex along the propagation axis by simple frequency modulation of the source.

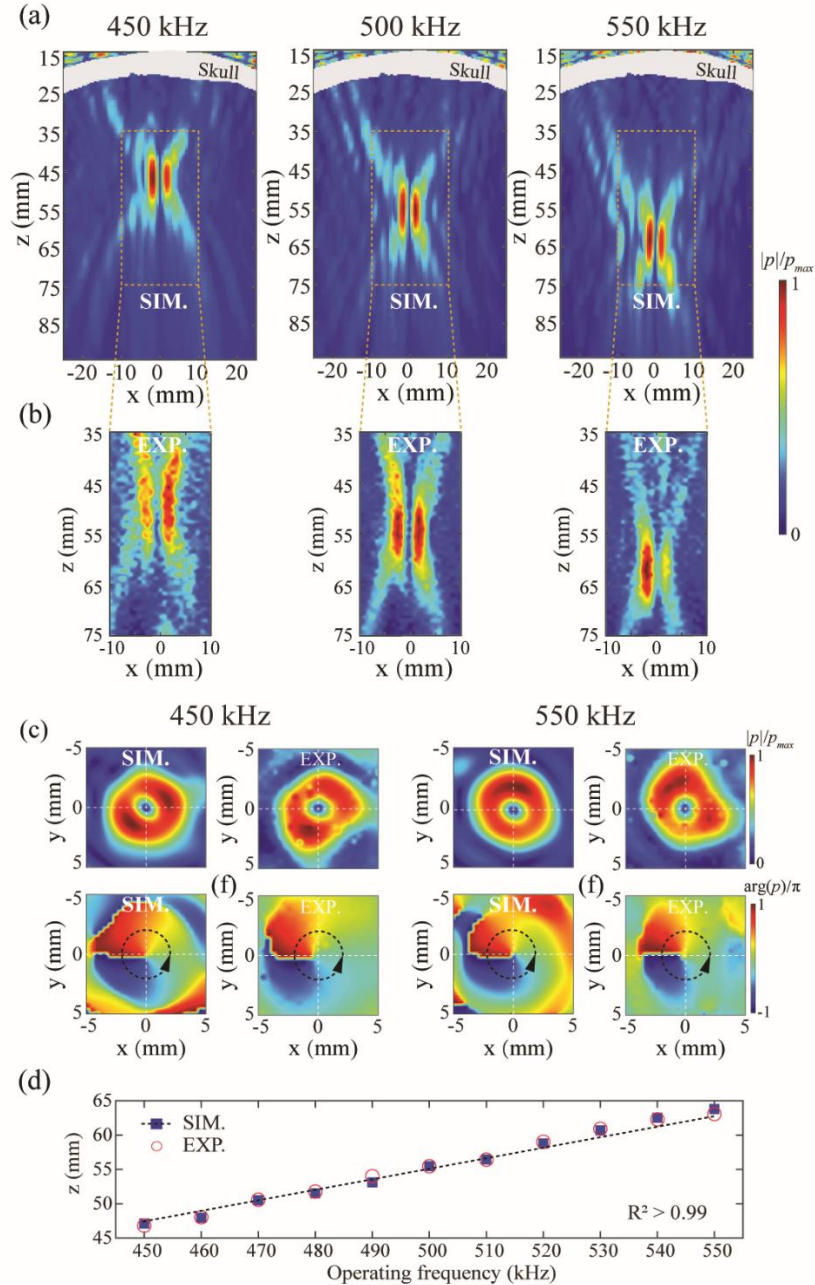

Fig. 4. The transcranial focusing depth of BAM is electronically steered by changing the operating frequency. (a-b) Simulated and experimentally measured ultrasound fields in the xz-plane at 450, 500, and 550 kHz, respectively. (c) Transverse cross section of the field at 450 and 550 kHz. (d) Simulated (solid squares) and experimentally measured (open circles) focal depths from 450 to 550 kHz.

## 4. Discussion

This study represents a significant step forward in using binary acoustic metasurfaces to manipulate ultrasound propagation through bone (specifically, the skull). We demonstrated that a thin, low-cost 3D-printed lens can correct skull-induced aberrations and generate complex waveforms like acoustic vortices inside the skull. Using a straightforward experimental setup (a single-element transducer with a snap-on lens), we achieved outcomes that traditionally required expensive phased arrays or adaptive optics. The BAM effectively encodes the intricate phase pattern obtained from time-reversal simulations, including both the focusing and vortex-generating components. Our findings highlight several key capabilities of the BAM approach: (1) It can produce focused vortex beams in homogeneous media with high fidelity (as shown in Fig. 2); (2) It can transcranially generate a focused vortex by pre-compensating for the skull's phase delays (as demonstrated in Fig. 3); and (3) It enables dynamic adjustment of the focal depth through frequency tuning, even with the skull in place (feature detailed in Fig. 4). Additionally, we numerically verified that the same BAM design can be extended to create higher-order vortices (with topological charges $l = 2, 3, 4$) through the skull, by simply changing the encoded vortex phase pattern (see Appendix, Fig. 5). Higher-order vortices may exert different radiation force patterns, which could be useful for advanced particle manipulation or therapeutic strategies.

Compared to the commonly used phased array technique, the proposed BAM for focused ultrasound vortex has several distinct advantages. Phased arrays have the flexibility to achieve multiple functionalities, such as beam aberration correction and beam steering; however, they are often expensive, bulky, and require complex electronic systems. The 3D-printed BAMs can also achieve beam aberration correction, and dynamic focusing along the wave propagation direction. Moreover, they are low-cost, easy-to-fabric, and do not require complex hardware. BAMs can potentially lower the barrier to the broad applications of tFUS by providing an affordable, simple-to-design and easy-to-use ultrasound device for customized usages. Although BAMs are not designed to replace phased arrays in tasks requiring dynamic 3D beam steering, they do offer flexible beamforming capabilities, with the added ability to dynamically steer the beam along the propagation direction. The binary design of BAM enables operation across a broadband frequency range and facilitates dynamic focusing capabilities, as corroborated by prior research[56,57,72]. This design allows the transmitted elements to maintain equal amplitude and phase shift across a wide frequency spectrum. In contrast, acoustic lenses crafted with gradient-phase designs face challenges in achieving these attributes.

Our approach also has some limitations. The binary phase approximation, while convenient, is a simplification of the ideal continuous phase profile. This can introduce some higher-order diffraction artifacts and slightly reduce peak intensity compared to a perfect phase plate. In practice, we found the binary encoding sufficient for a clean focus, but future designs might explore more than two phase levels or optimized coding patterns to further improve efficiency. Another limitation is that our experiments were conducted in a controlled laboratory setting with an ex vivo skull. In vivo conditions might introduce additional complexities such as tissue motion, skull temperature variations, or acoustic impedance mismatches at the transducer-skin interface. Our skull sample was also a partial section (skullcap); a full skull would require careful alignment of the lens to the patient's head, perhaps integrated into a helmet or coupling fixture. The BAM's material properties are another consideration – we used PLA, which worked well at our frequencies, but higher frequency applications or long-term use might benefit from materials with lower absorption or different acoustic impedance. Ongoing research is needed to test other 3D-printable materials and to assess the durability and reproducibility of the metasurfaces.

Despite these considerations, the ability to non-invasively focus and steer ultrasound through bone has clear clinical relevance. Transcranial applications could greatly benefit from this technology: for instance, non-invasive neuromodulation of the brain could be achieved without the need for a bulky transducer array, enabling wearable or portable therapeutic ultrasound devices. Targeted drug delivery to the brain (through blood-brain barrier opening) or blood clot lysis (sonothrombolysis) in stroke treatment are other examples where a patient-specific lens could be quickly manufactured and used with a single-element focused ultrasound system to treat through the skull. The acoustic radiation forces and torques generated by focused vortices may allow microscale manipulation of clots or cells in the cerebrovascular system, adding a new therapeutic dimension. In summary, while our work focused on a transcranial scenario, the underlying concept – a compact passive lens correcting for bone aberration – is general and could be applied to a variety of bone ultrasound challenges in medicine.

## 5. Conclusion

We have designed and fabricated binary acoustic metasurface (BAM) lenses for the transcranial steering of focused ultrasound vortices, demonstrating their efficacy in the context of ultrasonic propagation through the skull. These metasurfaces successfully correct skull-induced beam aberrations, enabling the formation of clear acoustic vortex fields within the brain region. By leveraging the binary phase design, we achieved dynamic focusing: the focal depth of the transcranial vortex beam could be electronically shifted by adjusting the driving frequency, without moving the transducer or lens. We also validated through simulations (and proof-of-concept experiments) that higher-order vortex beams can be generated transcranially using the same approach. The BAMs are inexpensive, simple to design using patient CT data, and easy to manufacture, making them attractive for personalized ultrasound therapy devices. This technology provides an affordable and effective solution for bone-related ultrasonic applications, especially transcranial focused ultrasound, where it could facilitate non-invasive neuromodulation, targeted brain drug delivery, or ultrasonic therapy through the skull. Our results lay the groundwork for further development of customized ultrasonic lenses that can improve the precision and accessibility of ultrasonic imaging and therapy in brain tissues though complex bone structure.

## 6. Funding

This work was supported by National Science Foundation of China (Grant No. 12404523), Guangdong Basic and Applied Basic Research Foundation (Grant No. 2023A1515110293), State Key Laboratory of Acoustics and Marine Information Chinese Academy of Sciences (Grant No. SKLA202408), and the Fundamental Research Funds for the Central Universities.

### APPENDIX: BAM for ultrasound vortex at different topological charge inside skull

To further investigate the flexibility and performance of BAM in transcranially generating higher-order topological charge acoustic vortices, we extended our investigation to vortices with higher topological charges ($l$ = 2, 3, and 4). Using the same design procedure, we modified the vortex-generating phase term to create these higher-order vortices while keeping the skull correction phase the same.

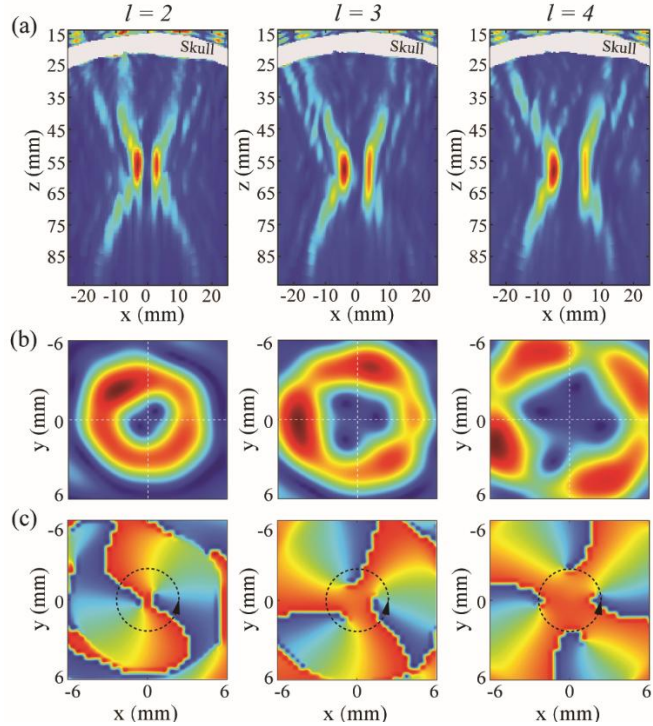

Fig. 5 Simulated fields representing topological charges ranging from $l$ = 2 to $l$ = 4 for transcranial focused ultrasound vortex. Displayed are the intensity distributions for the generated vortex beams along the *xz*-plane (a) and *xy*-plane (b), as well as the corresponding phase distribution along the *xy*-plane (c).

Figure 5 presents simulation results for $l$ = 2 to $l$ = 4 focused vortices inside the skull. The intensity distributions in the xz-plane (Fig. 5(a)) and the focal xy-plane (Fig. 5(b)) exhibit the expected expanding central null with increasing charge: as $l$ increases, the "doughnut" ring radius grows, and the on-axis null becomes wider. The phase maps (Fig. 5(c)) show $l$ intertwined $2\pi$ phase rotations around the center, confirming the charge of each vortex (e.g., two full 2π wraps for $l = 2$, etc.). These patterns highlight that the BAM can reliably produce higher-order vortices transcranially. Higher-order vortices carry more angular momentum and can exert different mechanical effects (e.g., $l = 2$ generates two intertwined helical wavefronts), which might be useful for specialized applications like multi-particle trapping or inducing shear waves in tissue. The success of these simulations indicates that our metasurface approach is not limited to single-charge vortices but can be generalized to a variety of complex ultrasonic beams for advanced therapeutic and diagnostic techniques involving bone.

## References

[1] X. Jiang, Y. Li, D. Ta, W. Wang, Ultrasonic sharp autofocusing with acoustic metasurface, Phys. Rev. B 102 (2020) 064308. https://doi.org/10.1103/PhysRevB.102.064308.

[2] X. Jiang, J.J. He, C.X. Zhang, H.L. Zhao, W.Q. Wang, D.A. Ta, C.W. Qiu, Three-dimensional ultrasound subwavelength arbitrary focusing with broadband sparse metalens, Sci. China Physics, Mech. Astron. 65 (2022) 1–7. https://doi.org/10.1007/s11433-021-1784-3.

[3] Q. Zhou, X. Ren, J. Huang, Z. Xu, X. Liu, Curved transport of microbubbles with compensated acoustic airy beams, Cell Reports Phys. Sci. 5 (2024) 101973. https://doi.org/10.1016/j.xcrp.2024.101973.

[4] Q. Zhou, Z. Xu, X. Liu, High efficiency acoustic Fresnel lens, J. Phys. D. Appl. Phys. 53 (2020). https://doi.org/10.1088/1361-6463/ab5878.

[5] Y. Li, K. Xu, Y. Li, F. Xu, D. Ta, W. Wang, Deep Learning Analysis of Ultrasonic Guided Waves for Cortical Bone Characterization, IEEE Trans. Ultrason. Ferroelectr. Freq. Control 3010 (2020) 1–1. https://doi.org/10.1109/tuffc.2020.3025546.

[6] X. Ren, Q. Zhou, J. Huang, Z. Xu, X. Liu, Holographic generation of arbitrary ultrasonic fields by simultaneous modulation of amplitude and phase, Ultrasonics 134 (2023) 107074. https://doi.org/10.1016/j.ultras.2023.107074.

[7] L. Meng, F. Cai, F. Li, W. Zhou, L. Niu, H. Zheng, Acoustic tweezers, J. Phys. D. Appl. Phys. 52 (2019). https://doi.org/10.1088/1361-6463/ab16b5.

[8] C. Ellouzi, A. Zabihi, F. Aghdasi, A. Kayes, M. Rivera, J. Zhong, A. Miri, C. Shen, Underwater double vortex generation using 3D printed acoustic lens and field multiplexing, APL Mater. 12 (2024). https://doi.org/10.1063/5.0201781.

[9] C.H. Fan, E. Huang, W.C. Lo, C.K. Yeh, Ultrasound-cavitation-enhanced drug delivery via microbubble clustering induced by acoustic vortex tweezers, Ultrason. Sonochem. (2025) 107273. https://doi.org/10.1016/j.ultsonch.2025.107273.

[10] Q. Zhou, J. Zhang, X. Ren, Z. Xu, X. Liu, Acoustic trapping of particles using a Chinese taiji lens, Ultrasonics 110 (2021) 106262. https://doi.org/10.1016/j.ultras.2020.106262.

[11] S. Guo, Z. Ya, P. Wu, M. Wan, A review on acoustic vortices: Generation, characterization, applications and perspectives, J. Appl. Phys. 132 (2022). https://doi.org/10.1063/5.0107785.

[12] A. Ashkin, J.M. Dziedzic, J.E. Bjorkholm, S. Chu, Observation of a single-beam gradient force optical trap for dielectric particles, Opt. Lett. 11 (1986) 288–290. https://doi.org/10.1364/ol.11.000288.

[13] B.T. Hefner, P.L. Marston, An acoustical helicoidal wave transducer with applications for the alignment of ultrasonic and underwater systems, J. Acoust. Soc. Am. 106 (1999) 3313–3316. https://doi.org/10.1121/1.428184.

[14] J.A. Davis, D.M. Cottrell, D. Sand, Abruptly autofocusing vortex beams, in: Front. Opt. 2012/Laser Sci. XXVIII, OSA, Washington, D.C., 2012: p. FW3A.58. https://doi.org/10.1364/FIO.2012.FW3A.58.

[15] C.J. Naify, C.A. Rohde, T.P. Martin, M. Nicholas, M.D. Guild, G.J. Orris, Generation of topologically diverse acoustic vortex beams using a compact metamaterial aperture, Appl. Phys. Lett. 108 (2016) 3–7. https://doi.org/10.1063/1.4953075.

[16] A. Marzo, M. Caleap, B.W. Drinkwater, Acoustic Virtual Vortices with Tunable Orbital Angular Momentum for Trapping of Mie Particles, Phys. Rev. Lett. 120 (2018) 44301. https://doi.org/10.1103/PhysRevLett.120.044301.

[17] A. Marzo, B.W. Drinkwater, Holographic acoustic tweezers, Proc. Natl. Acad. Sci. U. S. A. 116 (2019) 84–89. https://doi.org/10.1073/pnas.1813047115.

[18] X.D. Fan, Z. Zou, L. Zhang, Acoustic vortices in inhomogeneous media, Phys. Rev. Res. 1 (2019) 32014. https://doi.org/10.1103/PhysRevResearch.1.032014.

[19] H. Zhang, W. Zhang, Y. Liao, X. Zhou, J. Li, G. Hu, X. Zhang, Creation of acoustic vortex knots, Nat. Commun. 11 (2020) 3956. https://doi.org/10.1038/s41467-020-17744-x.

[20] Z.Y. Hong, J.F. Yin, B.W. Zhang, N. Yan, Vortex-field acoustic levitation in tubes, J. Appl. Phys. 128 (2020). https://doi.org/10.1063/5.0007554.

[21] S. Guo, Z. Ya, P. Wu, L. Zhang, M. Wan, Enhanced Sonothrombolysis Induced by High-Intensity Focused Acoustic Vortex, Ultrasound Med. Biol. 00 (2022) 1–11. https://doi.org/10.1016/j.ultrasmedbio.2022.05.021.

[22] M.A. Ghanem, A.D. Maxwell, Y.N. Wang, B.W. Cunitz, V.A. Khokhlova, O.A. Sapozhnikov, M.R. Bailey, Noninvasive acoustic manipulation of objects in a living body, Proc. Natl. Acad. Sci. U. S. A. 117 (2020) 16848–16855. https://doi.org/10.1073/pnas.2001779117.

[23] W.C. Lo, C.H. Fan, Y.J. Ho, C.W. Lin, C.K. Yeh, Tornado-inspired acoustic vortex tweezer for trapping and manipulating microbubbles, Proc. Natl. Acad. Sci. U. S. A. 118 (2021). https://doi.org/10.1073/PNAS.2023188118.

[24] V.M. Jooss, J.S. Bolten, J. Huwyler, D. Ahmed, In vivo acoustic manipulation of microparticles in zebrafish embryos, Sci. Adv. 8 (2022). https://doi.org/10.1126/sciadv.abm2785.

[25] Y. Yang, Y. Yang, D. Liu, Y. Wang, M. Lu, Q. Zhang, J. Huang, Y. Li, T. Ma, F. Yan, H. Zheng, In-vivo programmable acoustic manipulation of genetically engineered bacteria, Nat. Commun. 14 (2023) 3297. https://doi.org/10.1038/s41467-023-38814-w.

[26] D.-C. Chen, Q.-X. Zhou, X.-F. Zhu, Z. Xu, D.-J. Wu, Focused acoustic vortex by an artificial structure with two sets of discrete Archimedean spiral slits, Appl. Phys. Lett. 115 (2019). https://doi.org/10.1063/1.5108687.

[27] S. Ibsen, A. Tong, C. Schutt, S. Esener, S.H. Chalasani, Sonogenetics is a non-invasive approach to activating neurons in Caenorhabditis elegans, Nat. Commun. 6 (2015). https://doi.org/10.1038/ncomms9264.

[28] W. Legon, T.F. Sato, A. Opitz, J. Mueller, A. Barbour, A. Williams, W.J. Tyler, Transcranial focused ultrasound modulates the activity of primary somatosensory cortex in humans, Nat. Neurosci. 17 (2014) 322–329. https://doi.org/10.1038/nn.3620.

[29] K. Xu, Y. Yang, Z. Hu, Y. Yue, Y. Gong, J. Cui, J.P. Culver, M.R. Bruchas, H. Chen, TRPV1-mediated sonogenetic neuromodulation of motor cortex in freely moving mice, J. Neural Eng. 20 (2023) 016055. https://doi.org/10.1088/1741-2552/acbba0.

[30] Y. Yang, J. Yuan, R.L. Field, D. Ye, Z. Hu, K. Xu, L. Xu, Y. Gong, Y. Yue, A. V Kravitz, M.R. Bruchas, J. Cui, J.R. Brestoff, H. Chen, Induction of a torpor-like hypothermic and hypometabolic state in rodents by ultrasound, Nat. Metab. 5 (2023) 789–803. https://doi.org/10.1038/s42255-023-00804-z.

[31] Z. Hu, S. Chen, Y. Yang, Y. Gong, H. Chen, An Affordable and Easy-to-Use Focused Ultrasound Device for Noninvasive and High Precision Drug Delivery to the Mouse Brain, IEEE Trans. Biomed. Eng. 69 (2022) 2723–2732. https://doi.org/10.1109/TBME.2022.3150781.

[32] D. Ye, S. Chen, Y. Liu, C. Weixel, Z. Hu, J. Yuan, H. Chen, Mechanically manipulating glymphatic transport by ultrasound combined with microbubbles, Proc. Natl. Acad. Sci. 120 (2023) 2017. https://doi.org/10.1073/pnas.2212933120.

[33] D. Ye, J. Yuan, Y. Yang, Y. Yue, Z. Hu, S. Fadera, H. Chen, Incisionless targeted adeno-associated viral vector delivery to the brain by focused ultrasound-mediated intranasal administration, EBioMedicine 84 (2022) 104277. https://doi.org/10.1016/j.ebiom.2022.104277.

[34] S.H. Park, M.J. Kim, H.H. Jung, W.S. Chang, H.S. Choi, I. Rachmilevitch, E. Zadicario, J.W. Chang, Safety and feasibility of multiple blood-brain barrier disruptions for the treatment of glioblastoma in patients undergoing standard adjuvant chemotherapy, J. Neurosurg. 134 (2021) 475–483. https://doi.org/10.3171/2019.10.JNS192206.

[35] K. Beccaria, M. Canney, G. Bouchoux, C. Desseaux, J. Grill, A.B. Heimberger, A. Carpentier, Ultrasound-induced blood-brain barrier disruption for the treatment of gliomas and other primary CNS tumors, Cancer Lett. 479 (2020) 13–22. https://doi.org/10.1016/j.canlet.2020.02.013.

[36] Z. Hu, Y. Yang, D. Ye, S. Chen, Y. Gong, C. Chukwu, H. Chen, Targeted delivery of therapeutic agents to the mouse brain using a stereotactic-guided focused ultrasound device, STAR Protoc. 4 (2023) 102132. https://doi.org/10.1016/j.xpro.2023.102132.

[37] R. Martínez-Fernández, R. Rodríguez-Rojas, M. del Álamo, F. Hernández-Fernández, J.A. Pineda-Pardo, M. Dileone, F. Alonso-Frech, G. Foffani, I. Obeso, C. Gasca-Salas, E. de Luis-Pastor, L. Vela, J.A. Obeso, Focused ultrasound subthalamotomy in patients with asymmetric Parkinson's disease: a pilot study, Lancet Neurol. 17 (2018) 54–63. https://doi.org/10.1016/S1474-4422(17)30403-9.

[38] V. Krishna, P.S. Fishman, H.M. Eisenberg, M. Kaplitt, G. Baltuch, J.W. Chang, W.-C. Chang, R. Martinez Fernandez, M. Del Alamo, C.H. Halpern, P. Ghanouni, R. Eleopra, R. Cosgrove, J. Guridi, R. Gwinn, P. Khemani, A.M. Lozano, N. McDannold, A. Fasano, M. Constantinescu, I. Schlesinger, A. Dalvi, W.J. Elias, Trial of Globus Pallidus Focused Ultrasound Ablation in Parkinson's Disease., N. Engl. J. Med. 388 (2023) 683–693. https://doi.org/10.1056/NEJMoa2202721.

[39] R. Martínez-Fernández, J.U. Máñez-Miró, R. Rodríguez-Rojas, M. del Álamo, B.B. Shah, F. Hernández-Fernández, J.A. Pineda-Pardo, M.H.G. Monje, B. Fernández-Rodríguez, S.A. Sperling, D. Mata-Marín, P. Guida, F. Alonso-Frech, I. Obeso, C. Gasca-Salas, L. Vela-Desojo, W.J. Elias, J.A. Obeso, Randomized Trial of Focused Ultrasound Subthalamotomy for Parkinson's Disease, N. Engl. J. Med. 383 (2020) 2501–2513. https://doi.org/10.1056/nejmoa2016311.

[40] A.R. Rezai, M. Ranjan, P.F. D'Haese, M.W. Haut, J. Carpenter, U. Najib, R.I. Mehta, J.L. Chazen, Z. Zibly, J.R. Yates, S.L. Hodder, M. Kaplitt, Noninvasive hippocampal blood−brain barrier opening in Alzheimer's disease with focused ultrasound, Proc. Natl. Acad. Sci. U. S. A. 117 (2020) 9180–9182. https://doi.org/10.1073/pnas.2002571117.

[41] W.J. Elias, D. Huss, T. Voss, J. Loomba, M. Khaled, E. Zadicario, R.C. Frysinger, S.A. Sperling, S. Wylie, S.J. Monteith, J. Druzgal, B.B. Shah, M. Harrison, M. Wintermark, A pilot study of focused ultrasound thalamotomy for essential tremor, N. Engl. J. Med. 369 (2013) 640–648. https://doi.org/10.1056/NEJMoa1300962.

[42] L. Shu, H. Shen, Y. Zhou, Z. Li, Z. Hu, Harnessing the Power of Low-Intensity Focused Ultrasound Stimulation in Stroke: Modulating Immune Response, Stroke 0 (2025). https://doi.org/10.1161/STROKEAHA.125.052307.

[43] Z. Zhang, S. Ma, Z. Yin, J. Qiu, Z. Hu, G.-Y. Li, X.-Q. Feng, Y. Cao, Unveiling Hidden Features of Acoustic Radiation Forces in Soft Tissues Via Physics-Informed Neural Network-Based Full Shear Wave Inversion, J. Mech. Phys. Solids 205 (2025) 106326. https://doi.org/10.2139/ssrn.5220197.

[44] H. Gao, S. Ding, Z. Liu, J. Zhang, B. Li, Z. An, L. Wang, J. Jing, T. Liu, Z. Hu, Rapid Simulation Framework Integrating MRI-Derived Synthetic CT for Precise Transcranial Focused Ultrasound Targeting, Sci. CHINA Physics, Mech. Astron. (2025). https://doi.org/10.1007/s11433-025-2777-1.

[45] Z. Hu, Z. An, Y. Kong, G. Lian, X. Wang, The nonlinear S0 Lamb mode in a plate with a linearly-varying thickness, Ultrasonics 94 (2019) 102–108. https://doi.org/10.1016/j.ultras.2018.11.013.

[46] H. Montanaro, C. Pasquinelli, H.J. Lee, H. Kim, H.R. Siebner, N. Kuster, A. Thielscher, E. Neufeld, The impact of CT image parameters and skull heterogeneity modeling on the accuracy of transcranial focused ultrasound simulations, J. Neural Eng. 18 (2021). https://doi.org/10.1088/1741-2552/abf68d.

[47] S. Pichardo, V.W. Sin, K. Hynynen, Multi-frequency characterization of the speed of sound and attenuation coefficient for longitudinal transmission of freshly excised human skulls, Phys. Med. Biol. 56 (2011) 219–250. https://doi.org/10.1088/0031-9155/56/1/014.

[48] A.Y. Ammi, T.D. Mast, I.H. Huang, T.A. Abruzzo, C.C. Coussios, G.J. Shaw, C.K. Holland, Characterization of Ultrasound Propagation Through Ex-vivo Human Temporal Bone, Ultrasound Med. Biol. 34 (2008) 1578–1589. https://doi.org/10.1016/j.ultrasmedbio.2008.02.012.

[49] M. Baudoin, J.C. Gerbedoen, A. Riaud, O.B. Matar, N. Smagin, J.L. Thomas, Folding a focalized acoustical vortex on a flat holographic transducer: Miniaturized selective acoustical tweezers, Sci. Adv. 5 (2019) 1–7. https://doi.org/10.1126/sciadv.aav1967.

[50] A. Marzo, S.A. Seah, B.W. Drinkwater, D.R. Sahoo, B. Long, S. Subramanian, Holographic acoustic elements for manipulation of levitated objects, Nat. Commun. 6 (2015). https://doi.org/10.1038/ncomms9661.

[51] G.T. Clement, J. White, K. Hynynen, Investigation of a large-area phased array for focused ultrasound surgery through the skull, Phys. Med. Biol. 45 (2000) 1071–1083. https://doi.org/10.1088/0031-9155/45/4/319.

[52] J. White, G.T. Clement, K. Hynynen, Transcranial ultrasound focus reconstruction with phase and amplitude correction, IEEE Trans. Ultrason. Ferroelectr. Freq. Control 52 (2005) 1518–1522. https://doi.org/10.1109/TUFFC.2005.1516024.

[53] S. Jiménez-Gambín, N. Jiménez, F. Camarena, Transcranial Focusing of Ultrasonic Vortices by Acoustic Holograms, Phys. Rev. Appl. 14 (2020) 054070. https://doi.org/10.1103/PhysRevApplied.14.054070.

[54] G. Maimbourg, A. Houdouin, T. Deffieux, M. Tanter, J.-F. Aubry, 3D-printed adaptive acoustic lens as a disruptive technology for transcranial ultrasound therapy using single-element transducers, Phys. Med. Biol. 63 (2018) 025026. https://doi.org/10.1088/1361-6560/aaa037.

[55] Z. Hu, Y. Yang, L. Xu, Y. Hao, H. Chen, Binary acoustic metasurfaces for dynamic focusing of transcranial ultrasound, Front. Neurosci. 16 (2022) 1–9. https://doi.org/10.3389/fnins.2022.984953.

[56] B. Xie, K. Tang, H. Cheng, Z. Liu, S. Chen, J. Tian, Coding Acoustic Metasurfaces, Adv. Mater. 29 (2017) 1603507. https://doi.org/10.1002/adma.201603507.

[57] S. Tang, B. Ren, Y. Feng, J. Song, Y. Jiang, The generation of acoustic Airy beam with selective band based on binary metasurfaces: Customized on demand, Appl. Phys. Lett. 119 (2021) 071907. https://doi.org/10.1063/5.0060032.

[58] Z. Lin, X. Guo, J. Tu, Q. Ma, J. Wu, D. Zhang, Acoustic non-diffracting Airy beam, J. Appl. Phys. 117 (2015) 104503. https://doi.org/10.1063/1.4914295.

[59] N. Jiménez, V. Romero-García, V. Pagneux, J.-P. Groby, Quasiperfect absorption by subwavelength acoustic panels in transmission using accumulation of resonances due to slow sound, Phys. Rev. B 95 (2017) 014205. https://doi.org/10.1103/PhysRevB.95.014205.

[60] D. Tarrazó-Serrano, S. Pérez-López, P. Candelas, A. Uris, C. Rubio, Acoustic Focusing Enhancement In Fresnel Zone Plate Lenses, Sci. Rep. 9 (2019) 7067. https://doi.org/10.1038/s41598-019-43495-x.

[61] S. Jiménez-Gambín, N. Jiménez, J.M. Benlloch, F. Camarena, Holograms to Focus Arbitrary Ultrasonic Fields through the Skull, Phys. Rev. Appl. 12 (2019) 014016. https://doi.org/10.1103/PhysRevApplied.12.014016.

[62] B.E. Treeby, B.T. Cox, k-Wave: MATLAB toolbox for the simulation and reconstruction of photoacoustic wave fields, J. Biomed. Opt. 15 (2010) 021314. https://doi.org/10.1117/1.3360308.

[63] T.Y. Park, K.J. Pahk, H. Kim, Method to optimize the placement of a single-element transducer for transcranial focused ultrasound, Comput. Methods Programs Biomed. 179 (2019) 104982. https://doi.org/10.1016/j.cmpb.2019.104982.

[64] N. Wu, G. Shen, X. Qu, H. Wu, S. Qiao, E. Wang, Y. Chen, H. Wang, An efficient and accurate parallel hybrid acoustic signal correction method for transcranial ultrasound, Phys. Med. Biol. 65 (2020) 215019. https://doi.org/10.1088/1361-6560/abaa25.

[65] Z. Hu, L. Xu, C.-Y. Chien, Y. Yang, Y. Gong, D. Ye, C.P. Pacia, H. Chen, 3-D Transcranial Microbubble Cavitation Localization by Four Sensors, IEEE Trans. Ultrason. Ferroelectr. Freq. Control 68 (2021) 3336–3346. https://doi.org/10.1109/TUFFC.2021.3091950.

[66] U. Schneider, E. Pedroni, A. Lomax, The calibration of CT Hounsfield units for radiotherapy treatment planning, Phys. Med. Biol. 41 (1996) 111–124. https://doi.org/10.1088/0031-9155/41/1/009.

[67] T.D. Mast, Empirical relationships between acoustic parameters in human soft tissues, Acoust. Res. Lett. Online 1 (2000) 37–42. https://doi.org/10.1121/1.1336896.

[68] L. Marsac, D. Chauvet, R. La Greca, A.-L. Boch, K. Chaumoitre, M. Tanter, J.-F. Aubry, Ex vivo optimisation of a heterogeneous speed of sound model of the human skull for non-invasive transcranial focused ultrasound at 1 MHz, Int. J. Hyperth. 33 (2017) 635–645. https://doi.org/10.1080/02656736.2017.1295322.

[69] L. Xu, C.P. Pacia, Y. Gong, Z. Hu, C.-Y. Chien, L. Yang, H.M. Gach, Y. Hao, H. Comron, J. Huang, E.C. Leuthardt, H. Chen, Characterization of the Targeting Accuracy of a Neuronavigation-Guided Transcranial FUS System In Vitro, In Vivo, and In Silico, IEEE Trans. Biomed. Eng. 70 (2023) 1528–1538. https://doi.org/10.1109/TBME.2022.3221887.

[70] C. Constans, T. Deffieux, P. Pouget, M. Tanter, J.F. Aubry, A 200-1380-kHz quadrifrequency focused ultrasound transducer for neurostimulation in rodents and rrimates: Transcranial in vitro calibration and numerical study of the influence of skull cavity, IEEE Trans. Ultrason. Ferroelectr. Freq. Control 64 (2017) 717–724. https://doi.org/10.1109/TUFFC.2017.2651648.

[71] J.-F. Aubry, M. Tanter, M. Pernot, J.-L. Thomas, M. Fink, Experimental demonstration of noninvasive transskull adaptive focusing based on prior computed tomography scans, J. Acoust. Soc. Am. 113 (2003) 84–93. https://doi.org/10.1121/1.1529663.

[72] Z. Hu, Y. Yang, L. Xu, Y. Jing, H. Chen, Airy-Beam-Enabled Binary Acoustic Metasurfaces for Underwater Ultrasound-Beam Manipulation, Phys. Rev. Appl. 18 (2022) 024070. https://doi.org/10.1103/PhysRevApplied.18.024070.